# Advanced Image Enhancement Method for Distant Vessels and Structures in Capsule Endoscopy

Olivier Rukundo
Department of Computer Science, Norwegian University of Science and Technology
e-mail: olivier.rukundo@ntnu.no

Marius Pedersen
Department of Computer Science, Norwegian University of Science and Technology
e-mail: marius.pedersen@ntnu.no

Øistein Hovde
Department of Gastroenterology, Innlandet Hospital Trust, 2819 Gjøvik, Norway
e-mail: oistein.hovde@sykehuset-innlandet.no

*Abstract*—This paper proposes an advanced method for contrast enhancement of capsule endoscopic images, with the main objective to obtain sufficient information about the vessels and structures in more distant (or darker) parts of capsule endoscopic images. The proposed method (PM) combines two algorithms for the enhancement of darker and brighter areas of capsule endoscopic images, respectively. The half-unit weighted bilinear algorithm (HWB) proposed in our previous work is used to enhance darker areas according to the darker map content of its HSV's component V. Enhancement of brighter areas is achieved thanks to the novel thresholded weighted-bilinear algorithm (TWB) developed to avoid overexposure and enlargement of specular highlight spots while preserving the hue, in such areas. The TWB performs enhancement operations following a gradual increment of the brightness of the brighter map content of its HSV's component V. In other words, the TWB decreases its averaged-weights as the intensity content of the component V increases. Extensive experimental demonstrations were conducted, and based on evaluation of the reference and PM enhanced images, a gastroenterologist (ØH) concluded that the PM enhanced images were the best ones based on the information about the vessels, contrast in the images, and the view or visibility of the structures in more distant parts of the capsule endoscopy images.

## I. INTRODUCTION

In the effort to obtain more information about the vessels and structures, particularly, in the darker or distant parts of capsule endoscopic images, the image contrast enhancement is a way to go. There exist several categories and subcategories of contrast enhancement methods in the literature - for example, the Histogram Equalization (HE), Adaptive HE (AHE), and Contrast-Limited AHE (CLAHE) whose details are provided in [30],[31] as well as M1 a method proposed in [10], and M2 represents an enhancement method proposed in [9] - developed to deal generally with the poor contrast problems in color and grayscale images, which remain non-exhaustive and image dependent in their performance [1],[2],[3],[4],[5],[6],[7],[8]. Today, capsule endoscopy is among the newest research and application areas in medicine that caught interest of many researchers because of advantages capsule endoscopy provides over the traditional endoscopy in terms of comforting patients while exploring the entire gastrointestinal (GI) tract [10],[11],[12],[13],[23]. To clinically benefit from images obtained thanks to the capsule endoscope (CE), it is important to develop an advanced method that would deal carefully with the poor contrast problem caused generally by poor visibility conditions of the GI tract [14]. In this regard, a novel method using exclusively the bilinear interpolation algorithm has been proposed in [9] to deal with: (1) - the creation of artefacts leading to unnatural colors of the Histogram Equalization (HE) based methods - without the need for converting Red-Blue-Green (RGB) to another color space - [5] [6] and (2) - the disadvantages of the generalized overexposure problem of the method proposed previously in [10]. Although experimental demonstrations showed that the Half-unit weighted bilinear algorithm (HWB), proposed in [9], made considerable improvements over the method proposed in [10], (improvements that can also be seen/noticed in this paper's Figures 5-11 (a)), a gastroenterologist (ØH) could only rate highly 70 % of enhanced images presented in the [9]'s experimental demonstrations. Such a rating was due to over-enhancement of the neighborhoods of the brightest areas (i.e. specular areas) by the HWB and bad intensity transitions between darker and brighter areas, (as can be seen in this paper's Figures 5 - 11 (b)). Now, since the CE has the main light source - implanted in it, composed of a group of many Light Emitting Diodes - when such a light falls onto a GI surface tissue, some of the beams are reflected back straightaway -specular reflection - while the rest of the beams penetrate it before reflected (diffuse reflection) thus forming specular highlights on the capsule endoscopic images [25]. Although the sporadic presence of specular areas remains unavoidable, it is not a major one in this direction [24],[26], except enlarging them via over-enhancement of their neighborhoods. The advanced method - taking into account those possible enlargements and jagged transition of intensities between darker and brighter areas issues - has been proposed in this paper. The disadvantages of the proposed method (PM) is that it does not suppress cognitively (or under-enhance appropriately) specular highlight spots. It does not always perform very well or does not achieve the best scores with small sized images. Figure 1 shows the CE device and human GI tract.

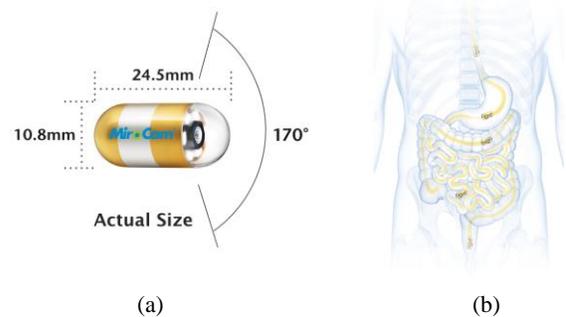

(a)           (b)

Figure 1: (a) Capsule endoscope (CE). (b) Gastrointestinal (GI) tract where the capsule passes. Images downloaded from the MiroCam® Capsule Endoscope from Medivators [22]. This Figure has also been used in [9].

This paper is organized as follows: The second part gives a brief introduction to the state of art key algorithm, dealing with darker areas, proposed in [9]. Part three gives the proposed method, as well as its summary. Experimental demonstrations, results, and evaluations by a gastroenterologist are provided in the fourth part. The conclusion is given in the fifth part.

## II. STATE OF THE ART

Half-unit weighted bilinear algorithm (HWB) is a novel RGB image enhancement strategy developed, in [9], to significantly remove all Histogram Equalization (HE) based artefacts and disadvantages[5], [6], rather than using complex enhancement strategies [15], [16]. A key point on which the HWB differs from the conventional bilinear algorithm (CWB) [17],[18],[19],[20] is that it uses a half-unit weighting strategy to calculate new pixel values for each overlapping four-pixel group in the destination matrix or image [9]. The mathematical expression on which the CWB is based is given in Equation 1. The $(r,c), (r,c+1)$, $(r+1,c)$ and $(r+1,c+1)$ are pixel ($P_n$) locations, on the pixel grid, as shown in Figure 2-b ($n$ is the number of four nearest neighbors).

$$CWB(r',c') = \sum_{n=1}^{4} P_n \times CW_n \quad (1)$$

where $CW_1 = (1-\Delta r) \times (1-\Delta c)$, $CW_2 = (\Delta r) \times (1-\Delta c)$, $CW_3 = (1-\Delta r) \times (\Delta c)$ and $CW_4 = (\Delta r) \times (\Delta c)$ represent the weighting functions in the CWB. $CWB(r',c')$ provides or presents interpolated values. Note that in Figure 2, $CWB(r',c')$ is also represented by $Y(r',c')$.

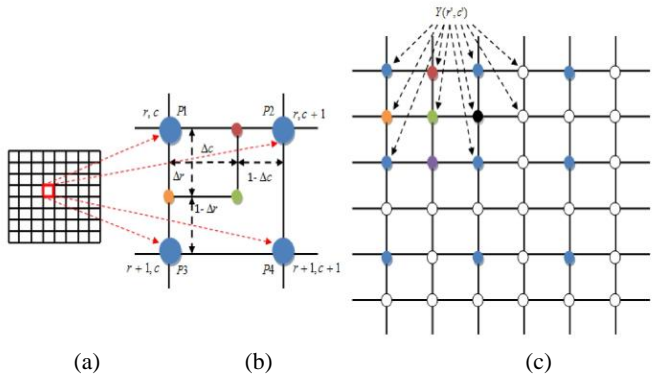

(a) (b) (c)

Figure 2: (a) represents the source pixel grid. (b) represents a four pixels group, (c) represents the destination pixel grid [9].

The mathematical expression for the HWB algorithm is given in Equation 2. Equation 2 is the result of applying a constant half-unit to weights $CW_n$ in Equation 1.

$$HWB(r',c') = \frac{\left(\sum_{n=1}^{4} P_n\right)}{HW} \quad (2)$$

where $HW = 2$ is the weighting function of the HWB. It is important to note that Equation 2 is the main function used to calculate the pixel values in the preliminary enhancement stage, as explained in [9].

## III. PROPOSED METHOD

The thresholded weighted-bilinear algorithm (TWB) is a novel algorithm, that operates from an empirical threshold value, $\partial$, and, the Hue-Saturation-Value (HSV) component V, developed and proposed, in this paper, to achieve the overall proposed method (PM). By developing TWB, the objective is to back the HWB algorithm and achieve an overall enhancement scheme that leads to a better visibility of distant capsule endoscopic images vessels and structures needed by gastroenterologist in their clinical diagnosis than what was previously achieved in [9]. The mathematical expression for the TWB's weighting function (TW) is given in Equation 3.

$$TW_{n(m)} = \frac{(CW_n + HW)}{H_m} \quad (3)$$

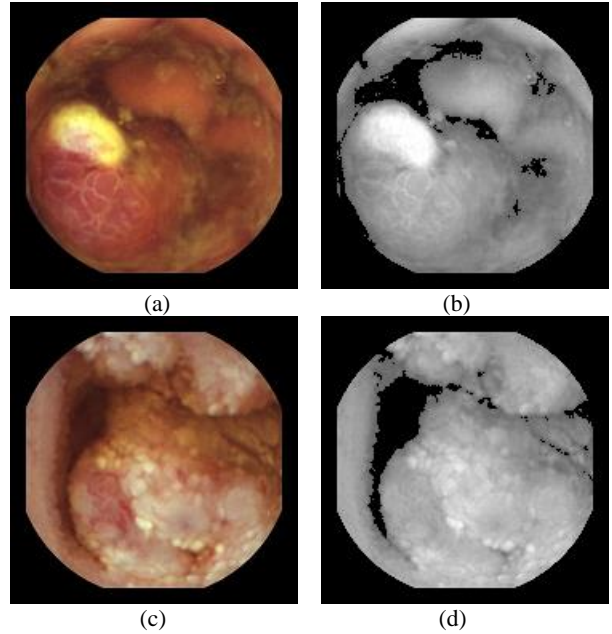

(a) (b)

(c) (d)

Figure 3: (a) and (c) are two test images from the capsule endoscopy database for medical decision support [21]. (b) and (d) show the darker and brighter maps in component V of (a) and (b), respectively. Here, the darker areas are defined by V $< \partial$. The brighter are defined by V $\geq \partial$.

$H_m$ is the denominator of the weighting function TW, whose mathematical expression is given in Equation 4.

$$H_m = \beta + \omega \quad (4)$$

where $\beta$ is the TW's denominator initial value whose optimal range, leading to a better linkage process between the darker and brighter areas (see Figure 3-(b)-(d)) has been experimentally located between 1.45 and 1.50; and $\omega$ is equal to the difference between consecutive $H_m$ denominators. Note that the experimental value for $\omega$ was found to be equal to or less than 0.025 so that the border between darker and brighter areas can become invisible. Here, $m$ is the number of weighting steps between the component V empirical threshold and maximum values, as shown in Equation 5. This number can be obtained by dividing the component V maximum value, $\varphi$, with the component V empirical threshold value $\partial$ (a default value for $\partial$ is equal to 0.4).

$$m = \frac{\varphi}{\partial} \qquad (5)$$

Note that $\varphi$ is equal to 1 in the component V. The mathematical expression for the TWB is given by Equation 6.

$$TWB(r',c')_m = \sum_{n=1}^{4} P_n \times TW_{n(m)} \qquad (6)$$

The PM's mathematical expression is given by Equation 7. The PM's Equation is a combination of the Equation 2 and Equation 6. The functioning of this combination is enabled by a set condition where it has to be verified whether the component V values are greater or less than the component V empirical threshold value $\partial$. If this condition is true (Yes), the matrix output of Equation 2 is added together with the reference matrix. If the condition is false (No), the matrix output of Equation 6 is added together with the reference matrix. The final matrix mapping, of all output matrices, constitutes the PM enhanced image. The summary of the PM is given in the following Figure 4. As can be seen/noticed the simplicity in the design of the PM is also another advantage computationally, although the processing time is not the main concern in this work.

$$PM(r',c') = \frac{\left(\begin{cases} HWB_{(r',c')} & if\ \eta<\partial \\ TWB_{(r',c')} & f\ \eta\geq\partial \end{cases}\right\} + I_{(r,c)}\right)}{2} \qquad (7)$$

where $I$ is the reference RGB image (often seen or treated as a poor contrast image), $\eta$ is any value of the V component, ranging between zero and $\varphi$.

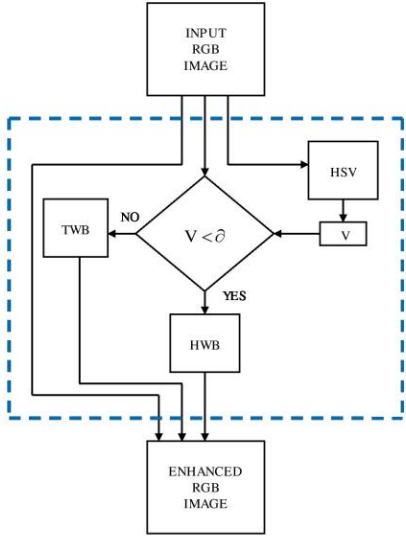

Figure 4: The proposed method (PM) summary

## V. EXPERIMENTS AND RESULTS

Experiments using standard image quality metrics, and the evaluation on better visibility of the vessels and structures in the PM enhanced images by a gastroenterologist, are presented here. The PM software has been implemented in MATLAB-R2017a. Image quality metrics used are structural similarity index (SSIM) and feature similarity index (FSIM).

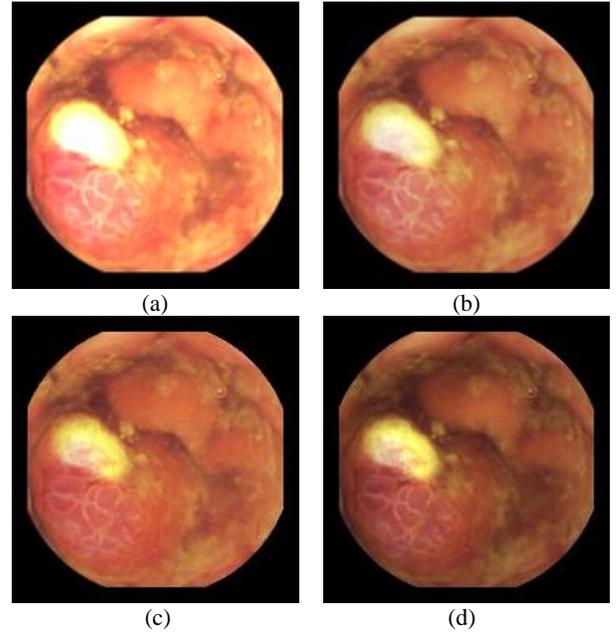

Figure 5: Image1, size 180 x 180, Image1(a) enhanced by M1, Image1(b) enhanced by M2, Image1(c) enhanced by PM, Image1(d) = reference image.

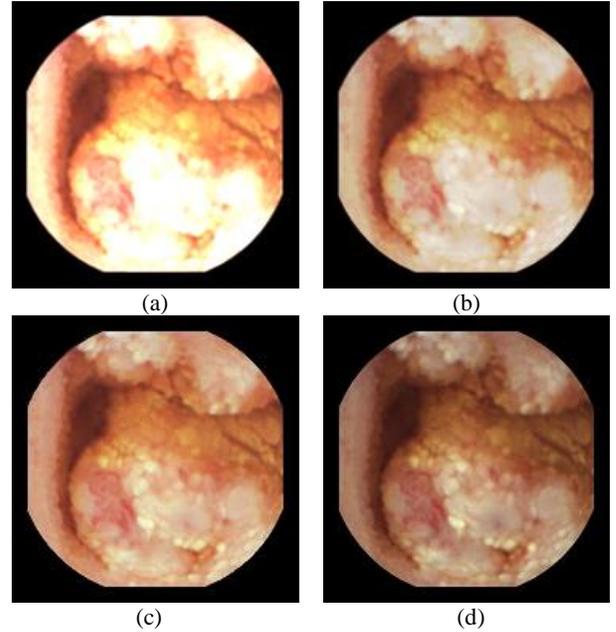

Figure 6: Image2, size 180 x 180, Image2(a) enhanced by M1, Image2(b) enhanced by M2, Image2(c) enhanced by PM, Image2(d) = reference image.

The reason is that, for capsule endoscopic applications where the pursuit for diagnostic quality is the main concern, metrics taking into account the image diagnostic structures and features (with reference to the reference image) are more appropriate than those who don't. It is important to note that there exist well-documented and widely available scientific works on such metrics in the literature [27],[28],[29]. Therefore, explanations, mathematical formulas, demonstrations, etc. of such metrics are not included in this paper. Also, metrics widely used in statistic of visual representation, such as contrast and intensity enhancement metrics, have been used to measure the contrast and intensity distortion in each RGB channel of the CE images. However, such methods normally apply on grayscale images. Therefore, their results presented were obtained based on processing each RGB channel separately. Blind/Referenceless Image Spatial QUality Evaluator (BRISQUE) which operates in the spatial domain, and according to [32] was the best performing metric for capsule images correlating with diagnostic quality, has been used to quantify possible losses of 'naturalness' in the enhanced images.

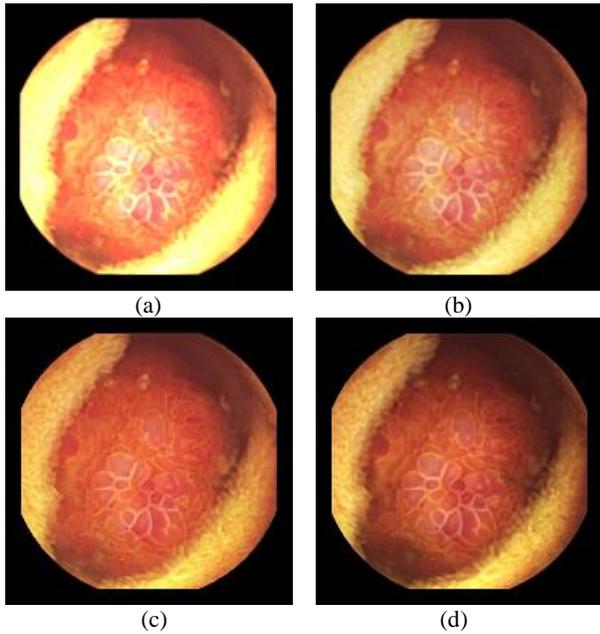

Figure 7: Image3, size 180 x 180, Image3(a) enhanced by M1, Image3(b) enhanced by M2, Image3(c) enhanced by PM, Image3(d) = reference image.

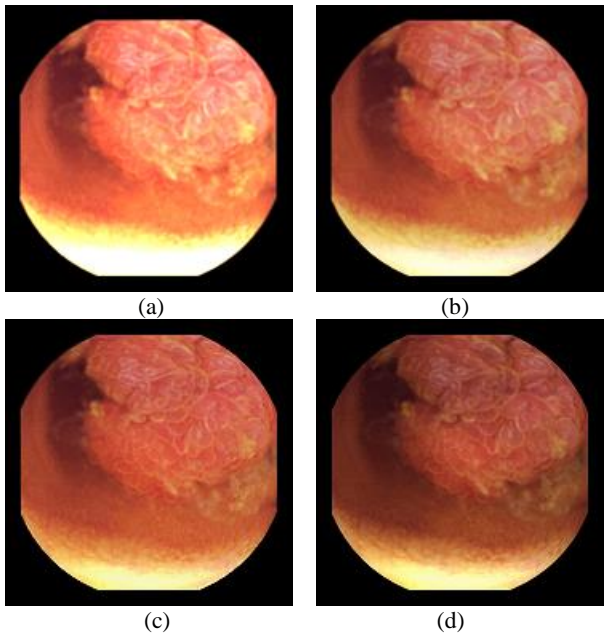

Figure 8: Image4, size 180 x 180, Image4(a) enhanced by M1, Image4(b) enhanced by M2, Image4(c) enhanced by PM, Image4(d) = reference image.

Capsule endoscopic images, downloaded from the capsule endoscopy database for medical decision support, have been used as test images[21].

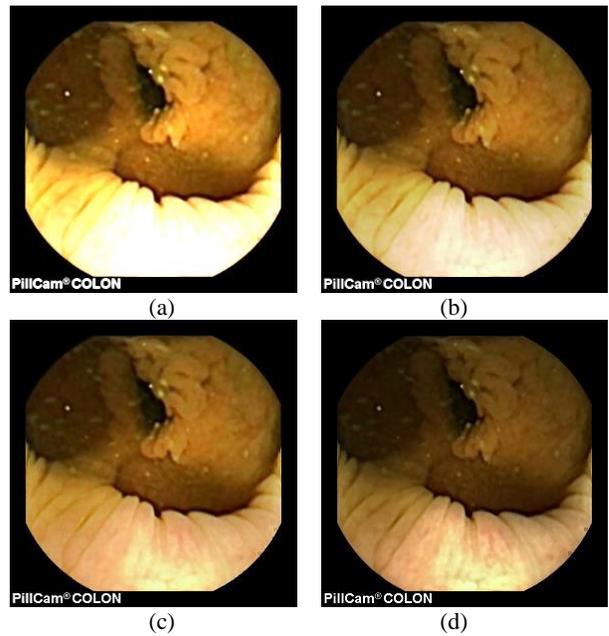

Figure 9: Image5, size 288 x 288, Image5(a) enhanced by M1, Image5(b) enhanced by M2, Image5(c) enhanced by PM, Image5(d) = reference image.

In Table 1 and Table 2, M1 represents the method proposed in [10], M2 represents the method proposed in [9], and PM represents the method proposed in this paper.

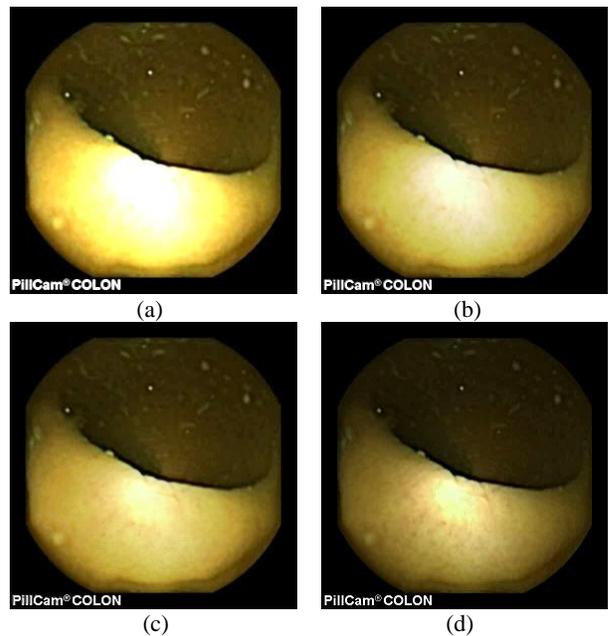

Figure 10: Image6, size 288 x 288, Image6(a) enhanced by M1, Image6(b) enhanced by M2, Image6(c) enhanced by PM, Image6(d) = reference image (with a highly visible specular highlight spot).

As can be seen, in images shown in Figure 5, Figure 6, Figure 7, Figure 8, Figure 9, Figure 10, Figure 11, only Figures 5 - 11 (c) are brighter, but not too bright, with preserved hue compared to Figures 5 - 11 (d). Furthermore, distant parts of capsule endoscopic images, in Figures 5 - 11 (c), are clearer or more visible than in the reference images. In this way, diagnostic details or information on the vessels and structures can be seen by a gastroenterologist better in Figures 5 - 11 (c) than in Figures 5 - 11 (a),(b),(d). In the ideal world, images obtained from capsule endoscopy would be perfect for a gastroenterologist in the sense of sharpness of image details, brilliant image colors, perfect image contrast and no artefacts. So far, there exist no such perfect capsule endoscopic images. However, based on evaluation of the reference, proposed, and even upscaled images - an additional experiment whose details are not included in this paper but which was conducted using Lanczos interpolation for three times (3X) upscaling purposes) - a gastroenterologist (ØH) concludes that the PM enhanced images were the best ones based on the information about the vessels, contrast in the images, and the view of the structures in the most distant parts of the images.

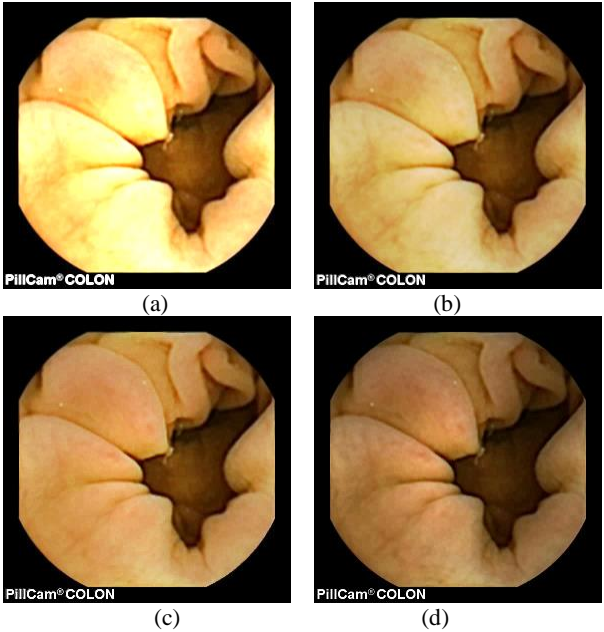

Figure 11: Image7, size 288 x 288, Image7(a) enhanced by M1, Image7(b) enhanced by M2, Image7(c) enhanced by PM, Image7(d) = reference image.

Table 1: Structural Similarity Index - SSIM

|  | *HE* | *AHE* | *CLAHE* | *M1* | *M2* | *PM* |
|---|---|---|---|---|---|---|
| *Image1* | 0.1237 | 0.6485 | 0.4201 | 0.7591 | 0.9061 | **0.9228** |
| *Image2* | 0.1708 | 0.5829 | 0.3924 | 0.7107 | 0.9024 | **0.9415** |
| *Image3* | 0.1035 | 0.6103 | 0.3675 | 0.8009 | 0.9253 | **0.9491** |
| *Image4* | 0.1041 | 0.6544 | 0.4318 | 0.7671 | 0.9142 | **0.9353** |
| *Image5* | 0.1078 | 0.6581 | 0.4156 | 0.6849 | 0.8885 | **0.9159** |
| *Image6* | 0.0820 | 0.6319 | 0.3673 | 0.7172 | 0.8929 | **0.9062** |
| *Image7* | 0.1003 | 0.6542 | 0.3985 | 0.7730 | 0.9133 | **0.9299** |

In some of the series the PM enhanced images were brighter, and, hence, it was easier to see the structures also in the distant parts of the images. In some of the series, the upscaled images were too blurred to give more information than the PM enhanced images, but most of the upscaled images gave more information than the normal sized ones. As mentioned earlier, details about Histogram Equalization (HE), Adaptive HE (AHE), and Contrast-Limited AHE (CLAHE) are provided in [30],[31]. As can be seen in both tables, the PM produced the highest SSIM and FSIMc values. The SSIM and FSIMc value closest or equal to one means generally the best quality because in that case, the similarity (to the reference image diagnostic quality structures and features) is almost maximum or maximum.

Table 2: Feature Similarity Index Color - FSIMc

|  | *HE* | *AHE* | *CLAHE* | *M1* | *M2* | *PM* |
|---|---|---|---|---|---|---|
| *Image1* | 0.7471 | 0.8014 | 0.7765 | 0.8685 | 0.9500 | **0.9660** |
| *Image2* | 0.8932 | 0.8144 | 0.8018 | 0.8383 | 0.9428 | **0.9741** |
| *Image3* | 0.8726 | 0.8282 | 0.8072 | 0.8133 | 0.9321 | **0.9690** |
| *Image4* | 0.8175 | 0.8123 | 0.7780 | 0.8673 | 0.9492 | **0.9728** |
| *Image5* | 0.8873 | 0.8397 | 0.8029 | 0.8674 | 0.9513 | **0.9618** |
| *Image6* | 0.7764 | 0.8716 | 0.8366 | 0.8734 | 0.9532 | **0.9579** |
| *Image7* | 0.8906 | 0.8553 | 0.8216 | 0.8945 | 0.9541 | **0.9646** |

Unlike in the M1 and M2 cases, the PM produced brighter images, but not too bright, and preserved the hue. It is important to note that Histogram Equalization (HE), Adaptive HE (AHE), and Contrast-Limited AHE (CLAHE) terribly change the hue of reference images, in RGB color space, as demonstrated in [5], [10]. Therefore, the capsule endoscopic images enhanced by such methods have not been included in this part. A part from widely known standard image quality metrics, in this paper, metrics widely used in statistic of visual representation have also been used, as started earlier, to measure the contrast and intensity distortion in each RGB channel of the CE images[23]. It is important to remind that, such methods normally apply on grayscale images. Therefore, results presented in Table 3 and Table 4 were obtained based on processing each RGB channel separately. It is also important to note that processing the intensities of each RGB channel is not the same as processing the intensities of a grayscale image. However, in the effort to find out how much enhancement methods affected each channel intensities, each channel was processed separately using those metrics. The results obtained proved that diagnostic quality cannot be correctly assessed based on the highest values of contrast or intensity enhancement in each channel's intensities, between the reference and enhanced images [23]. For example, in Table 3 the method proposed in [10] gave the highest values in terms of contrast enhancement (in all channels, for every image almost) but the corresponding images, shown in Figure 5 - 11 (a), showed that images produced by M1 are too bright and some image details are not visible compared to M2 and PM. Another example, in Table 4, showed that the HE gave the highest values in terms of intensity enhancement (in G and B channels, for every image) but as shown in [10], the output of the HE method does not give any usefully diagnostic information because damaging terribly the reference hue. However, if we analyze the statistics provided by these metrics in another way, for example, assuming that positive statistics means better quality, in this way the enhanced images and statistics by the PM have proven to be the most positively correlating to the reference image without overexposing or over-enhancing image details. Figure 12 presents the BRISQUE scores obtained. For image category whose size is equal to 288 x 288 x 3 (i.e. *image5*, *image6* and *image7*), the PM achieved generally the best scores. For image category whose size is equal to 188 x 188 x 3 (i.e. *image1*, *image2*, *image3* and *image4*), the PM achieved generally the second best scores. This suggests that the PM works better with larger images than with smaller images. However, the PM achieved the best scores compared to all enhancement methods mentioned.

## VI. CONCLUSION

Advanced enhancement method for vessels and structures in capsule endoscopic images has been proposed in this paper. The proposed method used mainly two HWB and TWB algorithms to deal with darker and brighter areas, respectively. It also used additional strategies to create a smooth intensity transition between such areas. The overall enhancement method achieved produced enhanced images with a moderately increase in brightness in darker/distant areas, and, that could preserve the hue of the reference images (without enlarging the specular highlight spots or over-enhancing their neighborhoods). Compared to the previous works, more details, especially in brighter areas could still be seen after the PM enhancement operations because the PM could avoid over-enhancing the neighborhood of the brighter areas. In this way, it was easier to see more details about the vessels and structures - for example in the pursuit of pre-cancerous or polypous tissues or even inflammations - in the PM enhanced images than in the reference images as well as in both M1 and M2 enhanced images. In the evaluation conducted together with a gastroenterologist (ØH) - a gastroenterologist concluded that PM enhanced images were the best ones based on the information about the vessels, the contrast in the images and the view of the structures in the most distant parts of capsule endoscopic images used. The usefulness of the PM enhanced images was also supported by statistics obtained using the SSIM and FSIMc metrics. Furthermore, in the effort to find out how much the PM affected each channel intensities, each channel was processed separately using contrast and intensity enhancement metrics. The first analysis demonstrated that the pursued diagnostic quality could not be correctly assessed based on the highest values of contrast or intensity enhancement in each channel's intensities, between the reference and enhanced images.

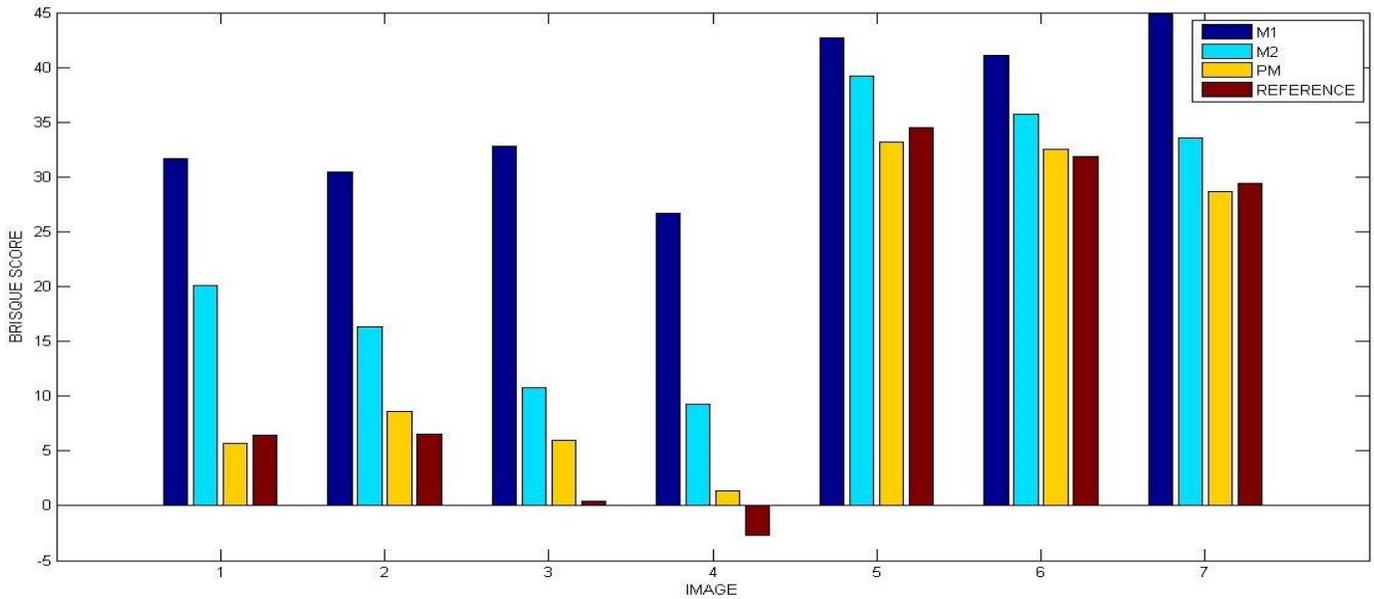

Figure 12: The comparison of M1, M2, PM and Reference Images based on BRISQUE scores

Table 3: Contrast Enhancement in RGB channels

|  | M1 | | | M2 | | | PM | | |
|---|---|---|---|---|---|---|---|---|---|
|  | *R* | *G* | *B* | *R* | *G* | *B* | *R* | *G* | *B* |
| *Image1* | **1.4871** | **1.7152** | 2.6847 | 0.6170 | 0.7056 | 1.1189 | 0.5971 | 0.3816 | 0.4361 |
| *Image2* | **1.1449** | **1.9602** | **2.6177** | 0.4896 | 0.8251 | 1.0918 | 0.5083 | 0.4953 | 0.4819 |
| *Image3* | **0.6263** | **1.5121** | 2.9125 | 0.2524 | 0.6325 | 1.2030 | 0.2408 | 0.1254 | 0.2000 |
| *Image4* | **0.9968** | 1.1078 | **2.1161** | 0.4093 | 0.4490 | 0.8893 | 0.3882 | 0.1263 | 0.1579 |
| *Image5* | **0.6197** | **1.0664** | **2.1690** | 0.2402 | 0.4448 | 0.9033 | 0.2174 | 0.1472 | 0.0823 |
| *Image6* | **0.8775** | **1.1147** | **1.9780** | 0.3650 | 0.4659 | 0.8229 | 0.3696 | 0.3790 | 0.2357 |
| *Image7* | **1.0862** | **2.3800** | **2.4949** | 0.4629 | 0.9928 | 1.0208 | 0.4755 | 0.5610 | 0.4777 |
|  | HE | | | AHE | | | CLAHE | | |
| *Image1* | -0.2728 | 0.7372 | **3.5268** | 0.1090 | 0.6680 | 1.6756 | -0.2641 | 0.2772 | 1.1922 |
| *Image2* | -0.3860 | 0.0399 | 0.6312 | 0.0222 | 0.5216 | 1.0866 | -0.3606 | -0.0143 | 0.3703 |
| *Image3* | -0.5415 | 0.0705 | **3.3189** | -0.1862 | 0.3917 | 2.4028 | -0.4900 | -0.0477 | 1.6584 |
| *Image4* | -0.4388 | **0.1519** | 1.7447 | -0.0520 | 0.3140 | 1.1327 | -0.3701 | -0.0410 | 0.6516 |
| *Image5* | -0.3852 | -0.1038 | 0.5886 | -0.1344 | -0.0435 | 0.3814 | -0.4766 | -0.3579 | 0.0057 |
| *Image6* | -0.2046 | -0.0317 | 1.1356 | -0.1869 | -0.0964 | 0.6501 | -0.5088 | -0.4126 | 0.1961 |
| *Image7* | -0.3498 | 0.2051 | 1.5053 | -0.0333 | 0.4124 | 1.5220 | -0.4192 | -0.0744 | 0.8891 |

Table 4: Intensity Enhancement in RGB channels

|  | M1 | | | M2 | | | PM | | |
|---|---|---|---|---|---|---|---|---|---|
|  | R | G | B | R | G | B | R | G | B |
| *Image1* | **0.7122** | 0.9117 | 1.0103 | 0.3577 | 0.4580 | 0.5084 | 0.3432 | 0.3306 | 0.3392 |
| *Image2* | **0.5706** | 0.8545 | 0.9769 | 0.2868 | 0.4290 | 0.4905 | 0.2841 | 0.3005 | 0.2945 |
| *Image3* | **0.4384** | 0.8199 | 1.0248 | 0.2205 | 0.4119 | 0.5154 | 0.2080 | 0.1886 | 0.2131 |
| *Image4* | **0.5781** | 0.8063 | 0.9597 | 0.2905 | 0.4052 | 0.4827 | 0.2762 | 0.2407 | 0.2514 |
| *Image5* | 0.5706 | 0.7112 | 0.9256 | 0.2871 | 0.3580 | 0.4670 | 0.2744 | 0.2539 | 0.1740 |
| *Image6* | 0.6501 | 0.7229 | 0.9192 | 0.3279 | 0.3648 | 0.4674 | 0.3242 | 0.3333 | 0.2722 |
| *Image7* | **0.5860** | 0.9307 | 1.0007 | 0.2948 | 0.4680 | 0.5051 | 0.2933 | 0.3257 | 0.3099 |
|  | HE | | | AHE | | | CLAHE | | |
| *Image1* | 0.4130 | **1.5860** | 2.9732 | 0.0055 | 0.3837 | 0.6665 | -0.0222 | 0.5710 | 1.0939 |
| *Image2* | 0.2875 | **0.8905** | 1.5424 | -0.0418 | 0.2090 | 0.4789 | -0.0958 | 0.2221 | 0.5715 |
| *Image3* | 0.2287 | **1.3171** | 3.0543 | -0.1207 | 0.2609 | 0.7773 | -0.1625 | 0.3879 | 1.1629 |
| *Image4* | 0.3425 | **1.5815** | 2.8684 | -0.0217 | 0.3595 | 0.6638 | -0.0416 | 0.5325 | 1.0731 |
| *Image5* | **0.6300** | 1.2110 | 3.6746 | 0.1006 | 0.2632 | 0.7072 | 0.1067 | 0.3879 | 1.3080 |
| *Image6* | **1.1621** | 1.5345 | 5.1613 | 0.1428 | 0.2312 | 0.9084 | 0.2708 | 0.4367 | 1.7947 |
| *Image7* | 0.3651 | **0.9644** | 2.7345 | 0.0288 | 0.2774 | 0.7700 | -0.0241 | 0.3060 | 1.0830 |

The second analysis suggested that the statistics provided by these metrics, in another way could mean better quality with reference to being closer to zero in a positive direction. And, enhanced images and statistics by the PM proved to be the most positively correlating to the reference image without overexposing or over-enhancing brighter areas neighborhoods and their neighborhoods and image textural details compared to M1 and M2 methods' outputs. Future work can be dedicated to the development of an innovative enhancement method enabling the desired gastroenterological sharpness of capsule endoscopic image details, color brilliantness, and artifact-free, as well as that, can lead to an under-enhancement of specular highlights spots since such spots hide the details in part of the image. On top of that since the PM gave better BRISQUE scores in 2 types of test images of the same size out of 3; and in only 1 type of test images of the same size out of 4 (smaller than the previous category size), an 'intelligent' or 'cognitive' method that would lead to the best visibility desired by gastroenterologists and scores in ALL types of test images sizes (in terms of BRISQUE, FSIM, etc.) will be dedicated to future efforts.

AUTHORS' CONTRIBUTION

OR conceived and designed the proposed method's algorithms, conducted experiments using standard image quality metrics as well as metrics widely used in statistic of visual representation and wrote the manuscript according to the expected standards in scientific publishing. ØH evaluated the PM enhanced images against reference images, in both normal and upscaled sizes, and validated the usefulness of PM enhanced images in terms of the information about the visibility of vessels, contrast in images, and structures for gastroenterological diagnosis. MP discussed the PM critically, gave feedback on the manuscript and contributed to the evaluation of the PM using the BRISQUE metric. All authors read and approved the final manuscript.

CONFLICT OF INTERESTS

Authors declare that there is no conflict of interest regarding the publication of this paper. In other words, authors confirm that the mentioned received funding in the acknowledgment section does not lead to any conflict of interests and that there is no any other possible conflict of interests regarding the publication of this paper.

ACKNOWLEDGMENT

The author would like to thank reviewers for their helpful comments. This research has been supported by the Research Council of Norway (project number: 260175, entitled: Upscaling based Image Enhancement for Video Capsule Endoscopy) through project number 247689: Image Quality enhancement in MEDical diagnosis, monitoring, and treatment - IQ-MED.